\documentclass[11pt, oneside]{article}   	
\usepackage{geometry}                		
\usepackage{graphicx}				
\usepackage{amssymb}
\usepackage{amsmath}
\usepackage{subfig}
\usepackage{caption}
\usepackage{comment}
\usepackage{amsthm,verbatim,amsfonts,amscd,float,physics}
\usepackage{appendix}

\usepackage[utf8]{inputenc}
\usepackage{xcolor}
\numberwithin{equation}{section}
\begin{document}
\title{Quantum Walks, Feynman Propagators and Graph Topology on an IBM Quantum Computer }
\author{Yuan Feng $^{(1)}$, Raffaele Miceli $^{(2)}$, Michael McGuigan $^{(3)}$\\
(1) Pasadena City College, (2) University of Pisa, (3) Brookhaven National Laboratory,}
\date{}
\maketitle
\begin{abstract}
Topological data analysis is a rapidly developing area of data science where one tries to discover topological patterns and connectivity in data sets to generate insight and knowledge discovery. In this project we use quantum walk algorithms to discover features of a data graph on which the walk takes place. This can be done faster on quantum computers where all paths can be explored using superposition. We begin with simple walks on a polygon and move up to graphs described by higher dimensional meshes. We use insight from the physics description of quantum walks defined in terms of probability amplitudes to go from one site on the graph to another distant site and show how this relates to the Feynman propagator or Kernel in the physics terminology. Our results from quantum computation using IBM’s Qiskit quantum computing software were in good agreement with those using classical computing methods.
\end{abstract}
\newpage

\section{Introduction - Classical vs. Quantum Walks}

Since their introduction in the 1990's \cite{PhysRevA.48.1687}, quantum walks have been studied extensively by many scientists, notably Childs, Farhi and Goldstone, \cite{2002quant.ph..9131C} \cite{2004PhRvA..70b2314C} \cite{2004PhRvA..70d2312C}\cite{foulger_quantum_2014} as a way to exponentially speed up certain graph search algorithms. In this paper we present a quick introduction to both classical and quantum walks and our efforts to implements quantum walks on small graphs in IBM's quantum simulator. Recent work implementing quantum walks on IBM quantum computers can be found in \cite{Acasiete_2020}.

Both classical and quantum walks occur on \textbf{graphs}, which are collections of \textbf{vertices} connected by \textbf{edges}. An important tool for studying graphs and a key element in graph walks is the \textbf{adjacency matrix}. A graph with $n$ vertices has an $n \times n$ adjacency matrix $A$, constructed as follows:

\begin{equation}
    A(i,j) = 
    \begin{cases}
    1, & i \neq j \ \text{and $\exists$ edge between vertices i and j} \\
    0, & \text{otherwise}
\end{cases}
\end{equation}

Below is an example visualization of a 6-vertex graph and its corresponding adjacency matrix:

\begin{figure}[H]
    \centering
    \begin{minipage}[b]{0.65\linewidth}
      \centering
      \includegraphics[width=\linewidth]{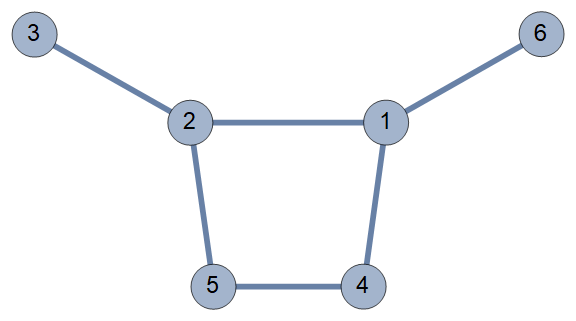}
    \end{minipage}
    \begin{minipage}[b]{0.35\linewidth}
      \centering
      \includegraphics[width=\linewidth]{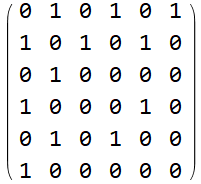}
    \end{minipage}
    \caption{A random graph with 6 vertices (left) and its adjacency matrix (right).}
    \label{graphadj6}
\end{figure}

\subsection{Classical Walks}

In a classical walk, the ``walker'' starts at a random vertex on the graph and has a certain probability of moving to any other vertex to which it is connected by an edge. These probabilities are controlled by the values in the adjacency matrix. Since in our case all the values are 0 or 1, the walker has an equal chance of moving to any vertex connected to its current vertex. The state of a walker on an $n$-vertex graph at time step $t$ is represented by a vector $\psi$ of length $n$, whose $j$-th entry counts the number of paths of length $t$ between the starting vertex and vertex $j$. $\psi(0)$ is a standard basis vector with a 1 at the index of the starting vertex and 0's everywhere else. We can evolve the state through repeated multiplication by the adjacency matrix:

\begin{equation}
    \psi(t) = A^t \psi(0)
\end{equation}

We can obtain the probabilities for the walker being state $j$ at time $t$ by dividing the corresponding entry in the state-vector by the total number of paths:

\begin{equation}
    P(j,t) = \frac{\psi_j(t)}{\sum_i \psi_i(t)}
\end{equation}

The evolution of a walker on the above graph, starting at vertex 1, is depicted in the plot below:

\begin{figure}[H]
    \centering
    \begin{minipage}[b]{0.7\linewidth}
      \centering
      \includegraphics[width=\linewidth]{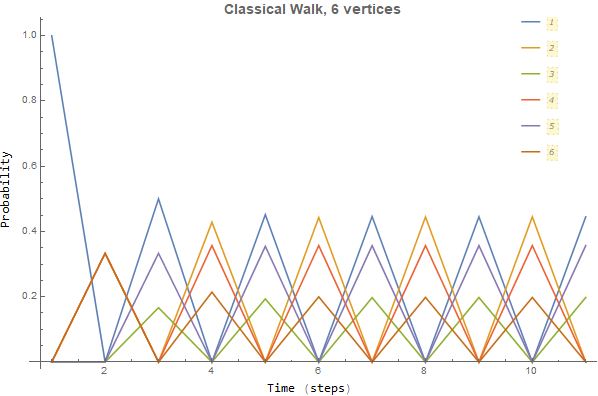}
    \end{minipage}
    \caption{Classical walk on the graph from figure \ref{graphadj6}}
    \label{cwalk6}
\end{figure}

In this case the quantum walk converges to periodic behavior after a small amount of steps, with the same state being swapped between two sets of vertices.

\subsection{Quantum Walks}

In a quantum walk, each vertex is treated as a possible state of a quantum system, in our case the qubits of a quantum computer. The walker can be in a superposition of any number of these states at any given time. The behavior of a quantum mechanical system over time can be calculated using the time-evolution operator, which is constructed from the Hamiltonian of the system. The time-evolution operator is an example of a \textbf{unitary operator} \cite{feynman2010quantum}, meaning that it preserves probability when it acts on a quantum state.

\begin{equation}
    \psi(t) = e^{-iHt}\psi(0)
\end{equation}

In the case of a quantum walk, the Hamiltonian $H$ is the adjacency matrix, the time $t$ varies continuously, and the state vector $\psi$ represents probability density:

\begin{equation}
    P(j,t) = |\psi_j(t)|^2 \quad , \quad \sum_i |\psi_i(t)|^2 = 1
\end{equation}

The starting state is the same as it was for the classical walk: a unit vector with a 1 in the index of the starting vertex and 0's everywhere else. Below is a quantum walk for the same starting conditions on the same graph from above:

\begin{figure}[H]
    \centering
    \begin{minipage}[b]{0.7\linewidth}
      \centering
      \includegraphics[width=\linewidth]{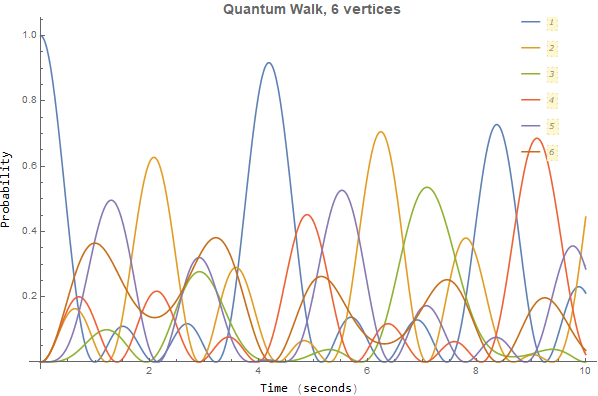}
    \end{minipage}
    \caption{Quantum walk on the graph from figure \ref{graphadj6}}
    \label{qwalk6}
\end{figure}

In the quantum walk, no periodic behavior is evident on this time scale, but we can still garner some information about the graph through qualitative inspection of the walk. For instance, the fact that the probabilities for vertices 2, 4 and 6 rise simultaneously at the start of the walk implies that they are directly connected to vertex 1. So if we were given an adjacency matrix for an unknown graph, we could ostensibly calculate the connectivity for each node by starting separate quantum walks at each vertex and observing the response of the others.

\section{Quantum Computing Methods}

The fundamental unit of a quantum computer is the qubit (quantum bit), a quantum two-state system. Qubits can be in either of the two states when measured, but can otherwise be in a \textbf{superposition} of those states. They can also be \textbf{entangled} with other qubits, making their state impossible to express independently. These two properties make quantum computers capable of running calculations in a way which is inaccessible to classical computers.
 
To run the quantum walks, we start by generating random graphs in Mathematica and exporting their adjacency matrices to text files. We then import them into a Python script which runs the quantum simulations using IBM's open-source QISkit software \cite{Qiskit}. The script uses the adjacency matrix to calculate the time-evolution operator for each time step we want, then translates the matrix operator into a quantum circuit composed of quantum gates which can be run on the quantum simulator. Each circuit is run many times in order to gather statistics.

\begin{figure}[H]
    \centering
    \begin{minipage}[b]{0.7\linewidth}
      \centering
      \includegraphics[width=\linewidth]{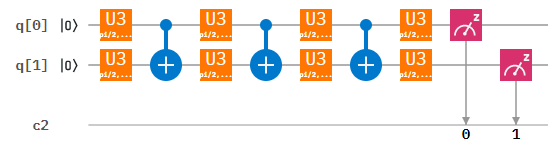}
    \end{minipage}
    \caption{Example circuit representing a 2-qubit unitary operator.}
    \label{2-qubit_unit}
\end{figure}

The Qiskit method used for the first part of the study, ''two\_qubit\_cnot\_decompose'', can only translate $2\times2$ (1-qubit) and $4\times4$ (2-qubit) unitary operators into quantum circuits. An example of a circuit representing a 2-qubit unitary operator can be seen in figure \ref{2-qubit_unit}.

\section{Trotter-Suzuki Simulations}

For larger Hamiltonians which require more qubits, we use Trotter-Suzuki Simulations to calculate the evolution of Hamiltonian. 

\begin{equation}
H=\sum_ja_jH_j
\end{equation}

In this function, each \(H_j\) has some certain form which makes the calculation of \(e^{-iH_jt}\) possible. 
The generalized form of Trotter formula as 

\begin{equation}e^{x\left(A+B\right)}=e^{p_1xA}e^{p_2xB}e^{p_3xA}e^{p_4xB}\cdots e^{p_MxB}+O\left(x^{m+1}\right)
\end{equation}

The set of parameters \(\lbrace p_1,p_2,\cdots,p_M\rbrace\) corresponding to order of \(x^{m+1}\). We use this approximation method to calculate the evolution of Hamiltonian. Given the Schrödinger equation,(we put \(\hbar=1\) here),

\begin{equation}i\frac{\partial}{\partial t}\psi\left(x,t\right)=H\psi\left(x,t\right)
\end{equation}

the Hamilton equation 

\begin{equation}
\frac{d}{dt} \binom{\vec{p} (t)}{\vec{q}(t) }=H\binom{\vec{p} (t)}{\vec{q}(t) }
\end{equation}

The time evolution operator of the Hamiltonian is described by the exponential operator
\begin{equation}
\binom{\vec{p} (t)}{\vec{q}(t) } =e^{iH}\binom{\vec{p} (0)}{\vec{q}(0)} 
\end{equation}

we use the Evolution of Hamiltonian (EOH) algorithm provided by Qiskit. We generate and export the Hamiltonian as before, using them as the input to construct the corresponding matrix operators, which could be converted to the Pauli spin operator representation. All tensor products of Pauli operators are 1-sparse. The local interactions can then be used to approximate the quantum system.  We then prepare the initial state on 3 qubits with probability 1 on a single vertex. Then we set up the quantum circuit for time evolution.

\begin{figure}[H]
    \centering
    \begin{minipage}[b]{0.7\linewidth}
      \centering
      \includegraphics[width=\linewidth]{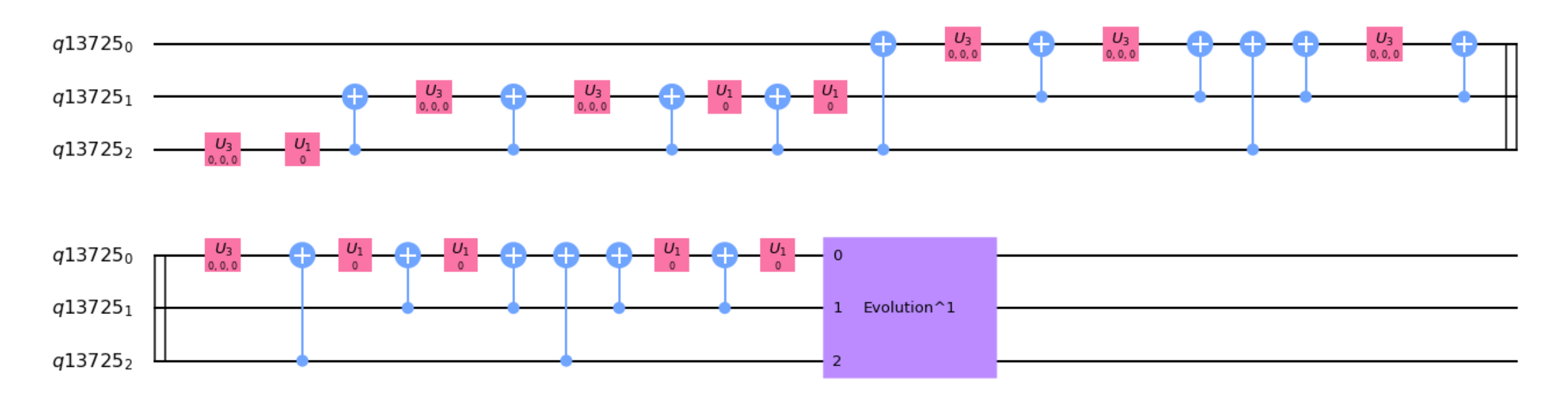}
    \end{minipage}
    \caption{Example circuit representing a 3-qubit EOH.}
    \label{3-qubit_unit}
\end{figure}

A larger number of time slices is needed to obtain an accurate result, especially for longer time intervals. We used the Suzuki-Trotter expansion mode for all the calculations, with an expansion order of 3. Since we use statevector simulator instead of the real quantum computer, we don’t need to measure all the qubits at the end of the circuit. Each run of the circuit returns a complex vector as a result, which defines the probability distribution of the property. To graph the results in a way that visualizes the evolution over time, we again perform multiple simulations at each time step.

\section{Quantum Computing Results}

We ran our procedure on many random graphs, and we include one example below. For the 4-vertex graph we split a time interval of 10 seconds into 200 steps and took 1000 measurements at each step. On a laptop running the Windows Subsystem for Linux (WSL) on Windows 10, these runs typically took less than 10 seconds to finish.

\begin{figure}[H]
    \centering
    \begin{minipage}[b]{0.5\linewidth}
      \centering
      \includegraphics[width=\linewidth]{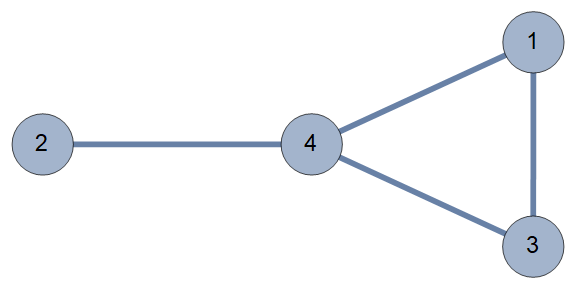}
    \end{minipage}
    \caption{A random graph with 4 vertices.}
    \label{graph4}
\end{figure}

\begin{figure}[H]
    \centering
    \begin{minipage}[b]{0.7\linewidth}
      \centering
      \includegraphics[width=\linewidth]{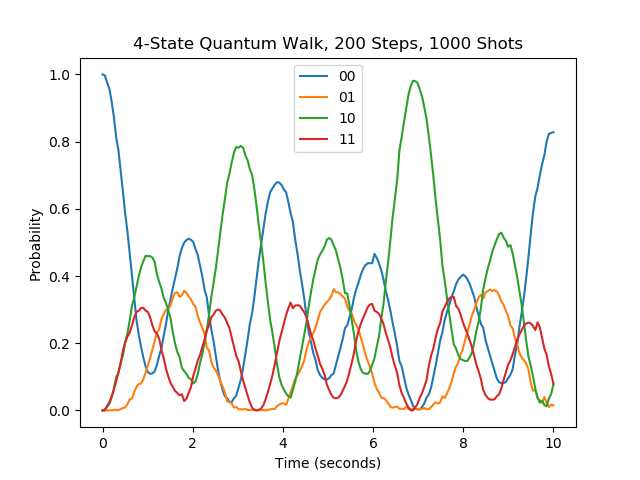}
    \end{minipage}
    \caption{Quantum walk on the graph from figure \ref{graph4} calculated using Qiskit.}
    \label{pwalk}
\end{figure}

\begin{figure}[H]
    \centering
    \begin{minipage}[b]{0.7\linewidth}
      \centering
      \includegraphics[width=\linewidth]{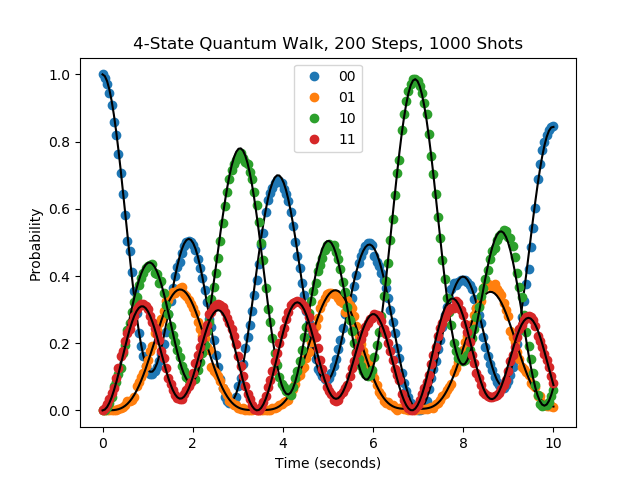}
    \end{minipage}
    \caption{Same quantum walk as above with the quantum calculations as dots and the exact probabilities as black lines to clarify the differences.}
    \label{pwalk2}
\end{figure}

\begin{figure}[H]
    \centering
    \begin{minipage}[b]{0.7\linewidth}
      \centering
      \includegraphics[width=\linewidth]{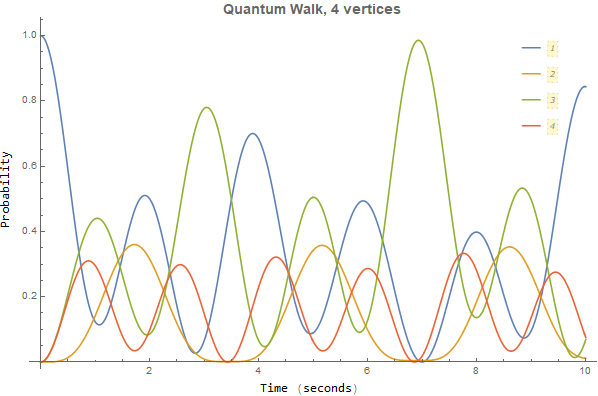}
    \end{minipage}
    \caption{Quantum walk on the graph from figure \ref{graph4} calculated using Mathematica.}
    \label{mwalk}
\end{figure}

Comparing the two quantum walks, we see that they match up almost exactly. The small discrepancies in the lower graph are due to the discrete and probabilistic nature of the calculation; while the walk in figure \ref{mwalk} was solved analytically, the walk in figure \ref{pwalk} was calculated by taking many measurements (shots) of the state at each time step and counting the number of times each state was measured. If we increase the number of shots, the lines should smooth out.

We can perform the same qualitative analysis on this quantum walk as we did before. We can see that the states corresponding to vertices 3 and 4 rise in probability simultaneously at the start of the walk, implying that those vertices are directly connected to vertex 1. Furthermore, we can see that vertex 4 rises less than vertex 3, indicating that probability might be ``leaking'' out of it into a different vertex, namely vertex 2.

With the EOH method, we tested more complicated cases with larger Hamiltonians. Here we include one such example involving an octagonal lattice and its corresponding $8\times8$ Hamiltonian. We start by preparing three qubits with the initial state \((1,0,0,0,0,0,0,0)\). For this run we used 10 time slices per second over 25 seconds, yielding 250 measurements.
\begin{figure}[H]
    \centering
    \begin{minipage}[b]{0.5\linewidth}
      \centering
      \includegraphics[width=\linewidth]{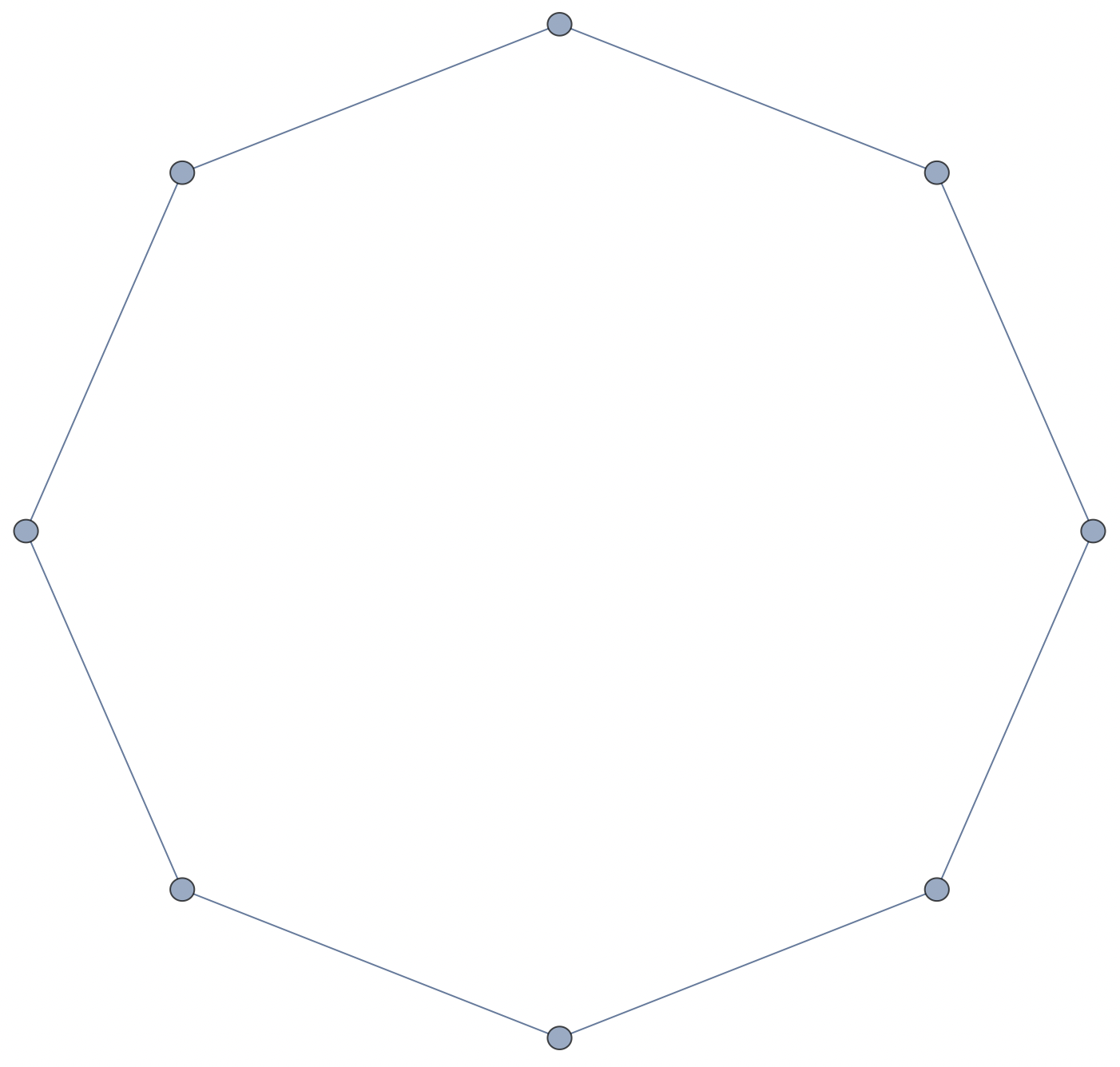}
    \end{minipage}
    \caption{Graph of Octagon}
    \label{3-qubit_unit}
\end{figure}

\begin{figure}[H]
    \centering
    \begin{minipage}[b]{0.7\linewidth}
      \centering
      \includegraphics[width=\linewidth]{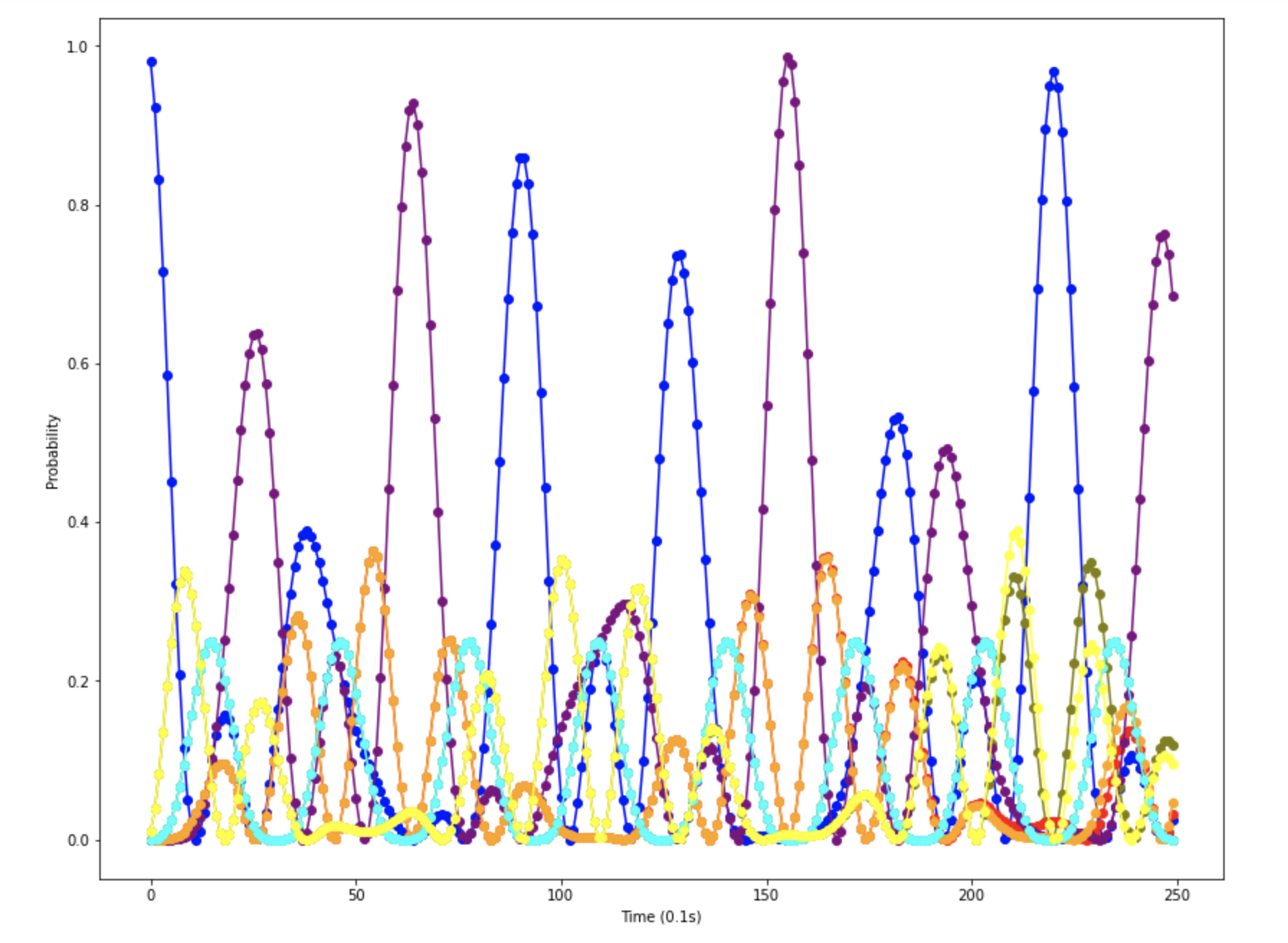}
    \end{minipage}
    \caption{Quantum walk on the graph from figure 10 calculated using Qiskit}
    \label{3-qubit_unit}
\end{figure}

\begin{figure}[H]
    \centering
    \begin{minipage}[b]{0.7\linewidth}
      \centering
      \includegraphics[width=\linewidth]{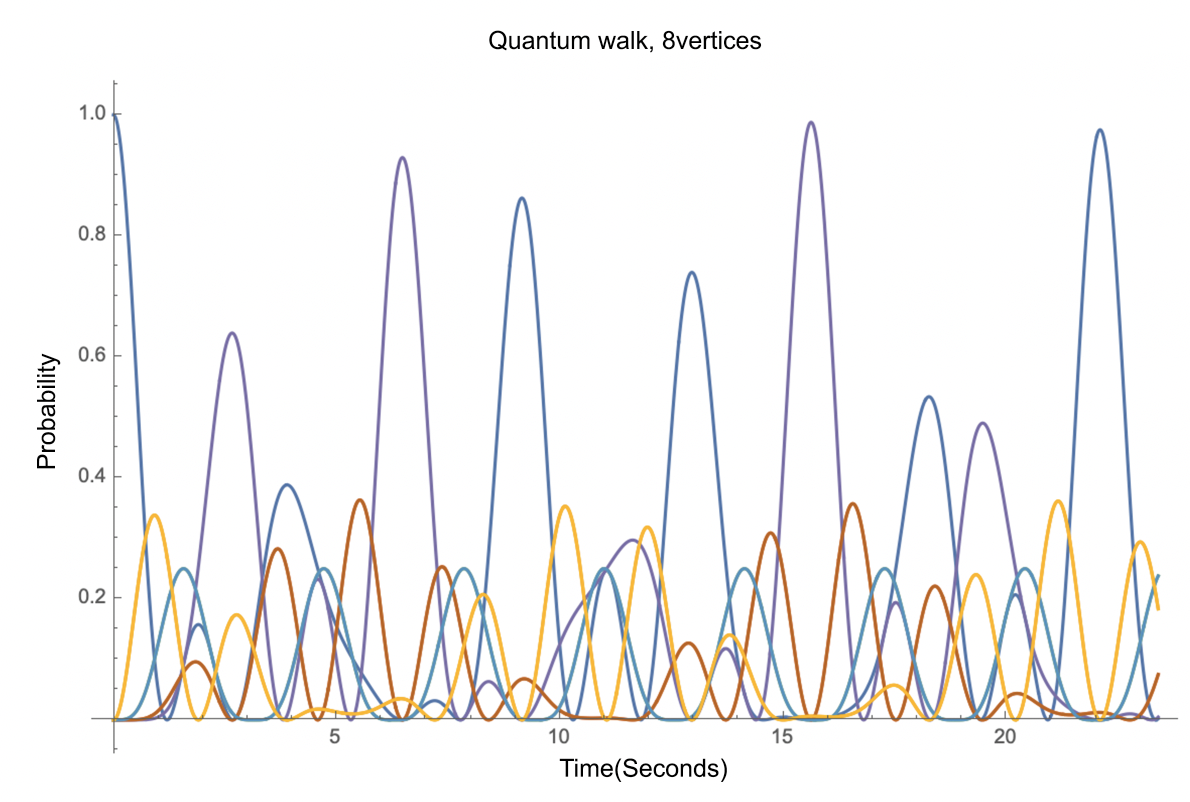}
    \end{minipage}
    \caption{Quantum walk on the graph from figure 10 calculated using Mathematica.}
    \label{3-qubit_unit}
\end{figure}

Comparing the two quantum walks, we see that they match up almost exactly. Similar to the case above, the small discrepancies in the lower graph are due to the discrete and probabilistic nature of the calculation.

\section{Relation of Quantum Walks to Feynman Propagators}

The Feynman propagator or Kernel represents the transition probability amplitude or Kernel function $K(x',x;t',t)$ for a particle whose position is known at position $x$ and time $t$ to be found at a new position $x'$ at time $t'$ \cite{Feynman:100771}. It is given in the Hamiltonian framework as :
\begin{equation}K(x',x;t' - t) = \left\langle {x',t'} \right|\left. {x,t} \right\rangle  = \left\langle {x'} \right|{e^{ - iH(t' - t)}}\left| x \right\rangle \end{equation}
which can be written in terms of energy eigenvalues $E_n$ and Eigenfunctions $\psi(x)$ as
\begin{equation}K(x,x';t) = \sum\limits_n {{e^{ - i{E_n}t}}{\psi_n ^ * }(x)\psi_n (x')} \end{equation}
or in terms of path integrals as:
\begin{equation}K(x',x;t' - t) = \int_{X(t) = x}^{X(t') - x'} {DX{e^{i\int_t^{t'} {d\bar tL(X(\bar t),\dot X(\bar t))} }}} \end{equation}
where $L$ is the Lagrangian of the system. For a free particle on the Real line the Hamiltonian is $H= \frac{P^2}{2m}$ with $P$ is the momentum operator and $m$ is the particles mass the kernel function was computed by Feynman and is given by:
\begin{equation}K(x,x';t) = {\left( {\frac{m}{{2\pi i\hbar t}}} \right)^{1/2}}{e^{ - m\frac{{{{\left( {x - x'} \right)}^2}}}{{2i\hbar t}}}}\end{equation}
For a particle on a discrete ring the Hamiltonian eigenvalues are:
\begin{equation}{E_n} = {\left( {\sin \left[ {\frac{{2\pi \left( {\frac{N}{2} + 1 - n} \right)}}{{2(N + 1)}}} \right]} \right)^2}, \{ n = 1, \ldots ,N\} \end{equation}
and the wave functions are:
\begin{equation}{\psi _n}(x) = \exp \left[ {i\frac{{2\pi \left( {\frac{N}{2} + 1 - n} \right)}}{{2(N + 1)}}x} \right]\end{equation}
For the discrete line interval the Hamiltonian eigenvalues are :
\begin{equation}{E_n} = {\left( {2\sin \frac{{\pi n}}{{2N}}} \right)^2, \{ n = 0, \ldots, N-1}\}\end{equation}
and wave functions are:
\begin{equation}{\psi _n}(x) = \cos \left[ {\frac{{\pi n}}{N}x} \right]\end{equation}
Plotting the Kernel for the discrete interval using formula we find perfect agreement with the quantum computation of the quantum walk as seen in Figure 13 and 14.
\begin{figure}[H]
    \centering
    \begin{minipage}[b]{0.7\linewidth}
      \centering
      \includegraphics[width=\linewidth]{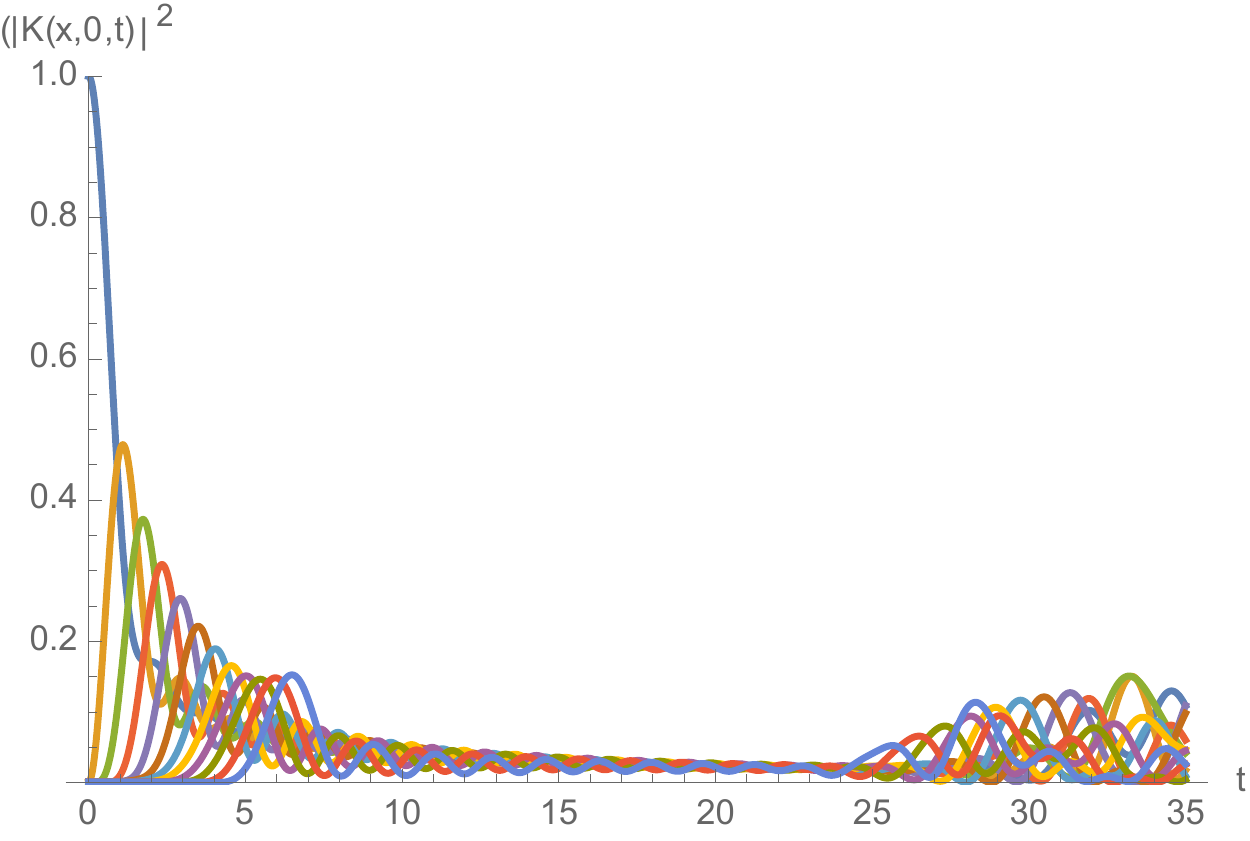}
    \end{minipage}
    \caption{Exact calculation of the Feynman propagator along an interval with 32 vertices.}
    \label{3-qubit_unit}
\end{figure}

\begin{figure}[H]
    \centering
    \begin{minipage}[b]{0.7\linewidth}
      \centering
      \includegraphics[width=\linewidth]{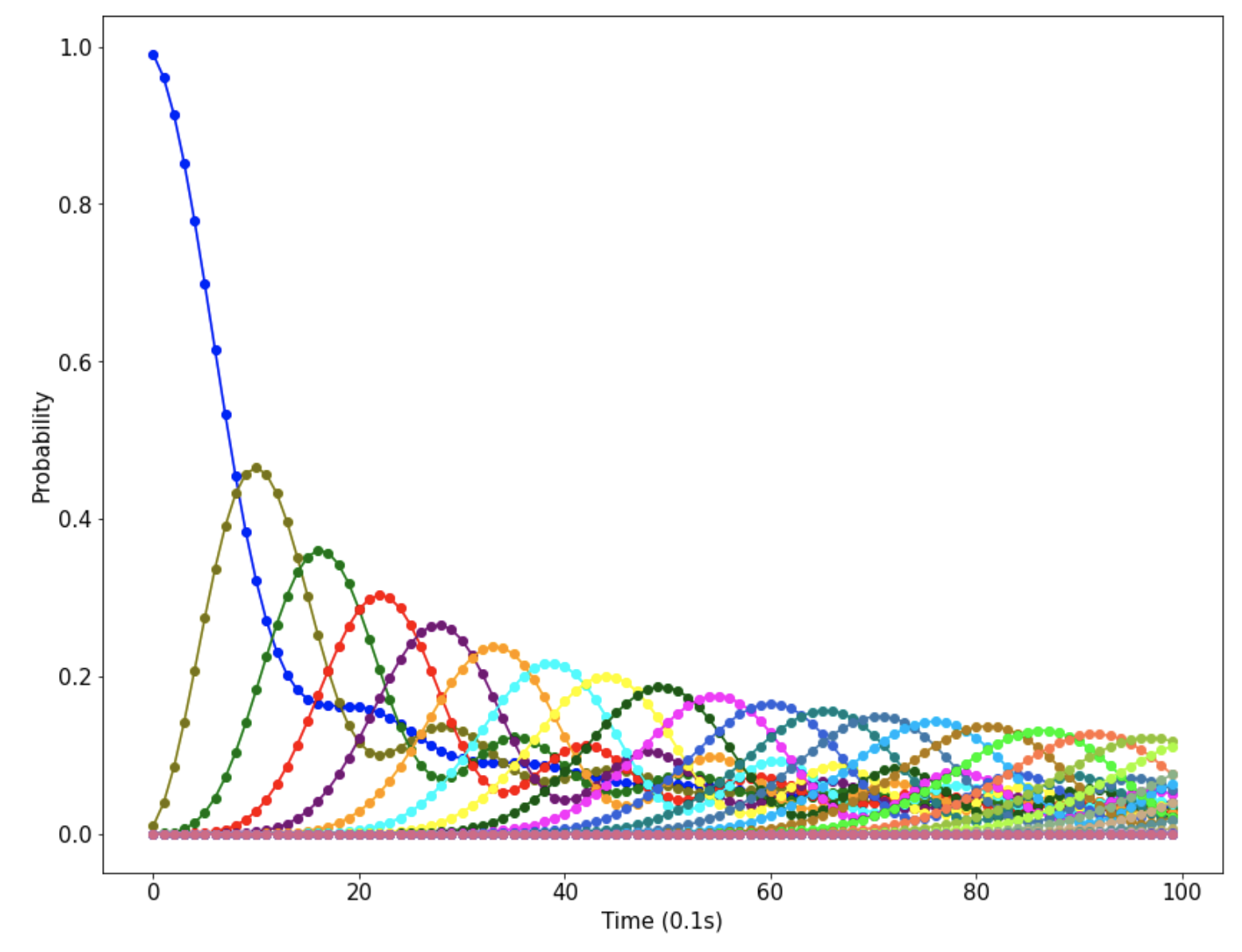}
    \end{minipage}
    \caption{Quantum computation of the quantum walk along an interval with 32 vertices.}
    \label{3-qubit_unit}
\end{figure}

\section{Relation of quantum walks to Grover search}

Grover's algorithm \cite{Grover:1996rk}\cite{Guillet:2019dws}is a quantum algorithm that can be used to find the target from an unsorted database of N items using a black box function with \(O\left(\sqrt{N}\right)\) evaluations of the function, which is more efficient than the classical algorithm with \(O\left(N\right)\). In Grover's algorithm, we define oracles by adding a negative phase to the solution states for any state \(\vert x \rangle \) in the computational basis. 
\begin{equation}U_w\vert x \rangle=\left(-1\right)^{f\left(x\right)} \vert x \rangle\end{equation}
We take the 2 qubits case as an example, setting target as \(\vert 11\rangle \) for \(N = 4\). 
\begin{equation}U_\omega \vert s \rangle = U_\omega \frac{1}{2} \left( \vert 00 \rangle + \vert 01 \rangle + \vert 10 \rangle + \vert 11 \rangle \right) = \frac{1}{2} \left( \vert 00 \rangle + \vert 01 \rangle + \vert 10 \rangle - \vert 11 \rangle \right)\end{equation}
\begin{equation}U_w=\begin{bmatrix}
1 & 0 & 0 & 0\\
0 & 1 & 0 & 0\\ 
0 & 0 & 1 & 0\\ 
0 & 0 & 0 & -1\\ 
\end{bmatrix}\end{equation}
For the spatial search case, we also set the target as \(\vert 11\rangle \) on a graph with 4 vertices. We have the oracle Hamiltonian as \(H_\omega = -\vert \omega \rangle \langle \omega \vert \), which could also be obtained by the standard Grover oracle. We start with a uniform superposition over all the vertices of the graph given by \(\vert s \rangle =\frac{1}{\sqrt{N}}\sum_j \vert j \rangle \). The time-independent Hamiltonian is given as
\begin{equation}H=-\gamma L+H_{\omega}\end{equation}
with the Laplacian as \(L=A-D\).

\begin{figure}[H]
    \centering
    \begin{minipage}[b]{0.7\linewidth}
      \centering
      \includegraphics[width=\linewidth]{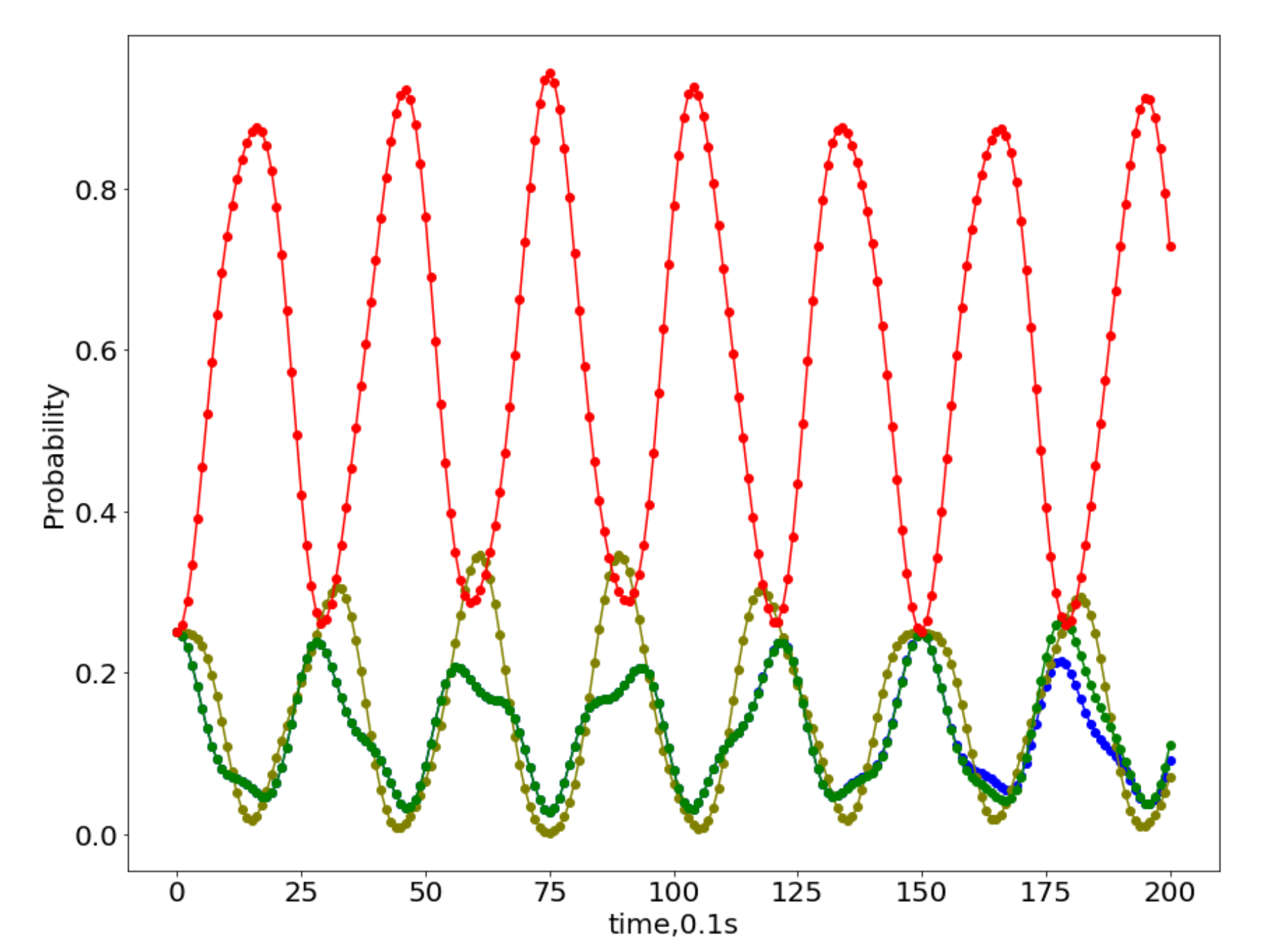}
    \end{minipage}
    \caption{spatial search for the target vertex}
    \label{3-qubit_unit}
\end{figure}

\section{Conclusion}

We created a workflow to calculate a quantum walk on any  graph using a quantum computer. A lot of information about a graph can be gleaned from a visual inspection of the quantum walk plot. For instance, the number of distinct lines in the plot reveals how many different classes of vertices there are in the graph relative to the starting vertex. Furthermore, if any other vertex ever reaches a probability arbitrarily close to 1, it is likely that it has connectivity similar or identical to the starting vertex. This concept of pretty good state transfer \cite{couvertier2018quantum} can be used to gain information about the symmetry of a graph. In the Appendix we give more examples of perfect state transfer. In this case there will eventually be a probability of exactly 1 of measuring a vertex besides the starting vertex. In cyclic graphs with an even number of vertices this always happens for the vertex opposite the starting vertex.


The relation of quantum walks to path integrals opens up new areas of applications to the simulation of quantum physical systems \cite{Feynman:100771} and quantum finance \cite{Egger_2020}\cite{bouland2020prospects}. The relation of quantum walks to search problems opens up new opportunities in large scale data analysis and protein folding using quantum computers \cite{casares2021qfold}. Quantum walks could potentially be applied to study the structure of systems which transition between different states. One such application is the problem of studying the decay chains of radioactive elements. The different isotopes in the chain would be the vertices of the graph, and in this case directed edges would be used with weights that encode the decay probabilities and background conditions. Another interesting problem might be to reverse our process, figuring out how to obtain the corresponding adjacency matrix or graph from the time evolution operator. This problem could rephrased as "Can one hear the topology of a quantum drum?" as a generalization of the classic math problem \cite{Kac}. This inverse method is similar to the experimental determination of nano-structures from X-ray scattering data \cite{162351}. 

\section*{Acknowledgements}
This material is based upon work supported in part by the U.S. Department of Energy, Office of Science, National Quantum Information Science Research Centers, Co-design Center for Quantum Advantage (C2QA) under contract number DE-SC0012704. This project was also supported in part by the U.S. Department of Energy, Office of Science, Office of Workforce Development for Teachers and Scientists (WDTS). We thank Chen-Fu Chiang for productive discussions on Quantum Walks.
\bibliographystyle{ieeetr}
\bibliography{references}

\newpage

\section{Appendix A - Some more 4-vertex quantum walks}

\begin{figure}[H]
    \centering
    \begin{minipage}{0.3\linewidth}
      \centering
      \includegraphics[width=\linewidth]{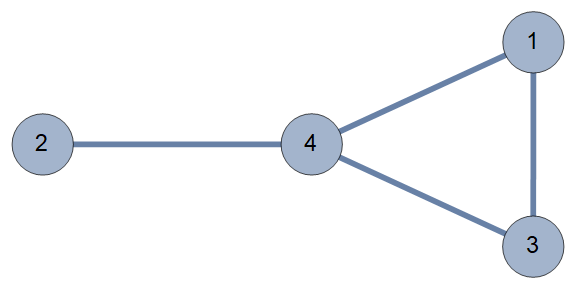}
    \end{minipage}
    \begin{minipage}{0.7\linewidth}
      \centering
      \includegraphics[width=\linewidth]{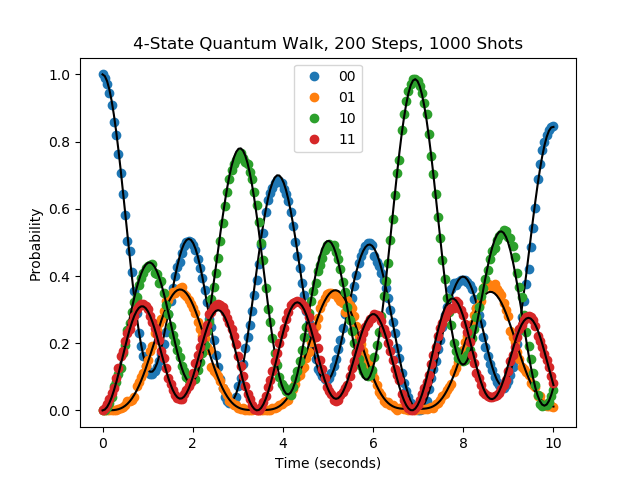}
    \end{minipage}
\end{figure}

\begin{figure}[H]
    \centering
    \begin{minipage}{0.3\linewidth}
      \centering
      \includegraphics[width=\linewidth]{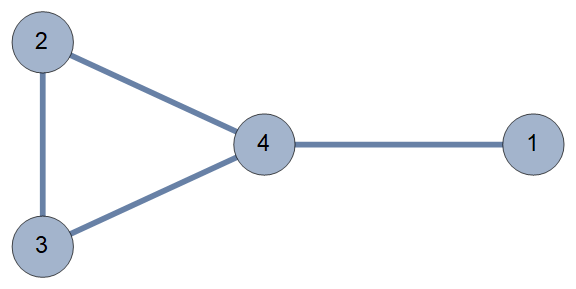}
    \end{minipage}
    \begin{minipage}{0.7\linewidth}
      \centering
      \includegraphics[width=\linewidth]{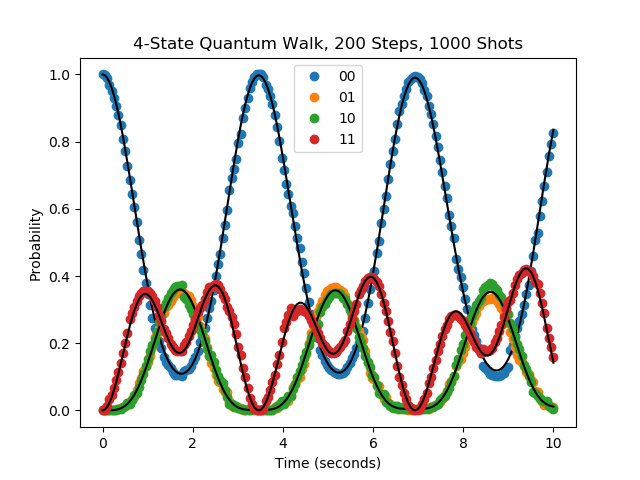}
    \end{minipage}
\end{figure}

\begin{figure}[H]
    \centering
    \begin{minipage}{0.3\linewidth}
      \centering
      \includegraphics[width=\linewidth]{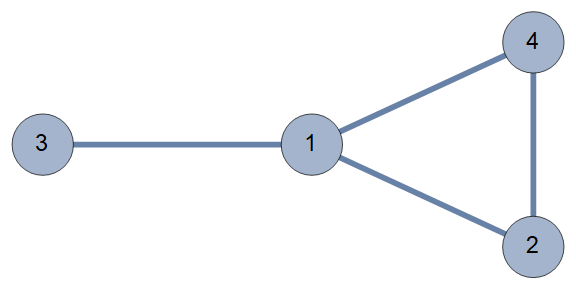}
    \end{minipage}
    \begin{minipage}{0.7\linewidth}
      \centering
      \includegraphics[width=\linewidth]{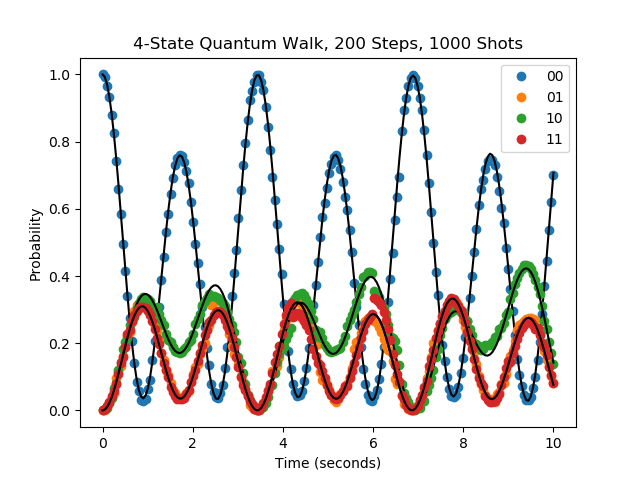}
    \end{minipage}
\end{figure}

\begin{figure}[H]
    \centering
    \begin{minipage}{0.3\linewidth}
      \centering
      \includegraphics[width=\linewidth]{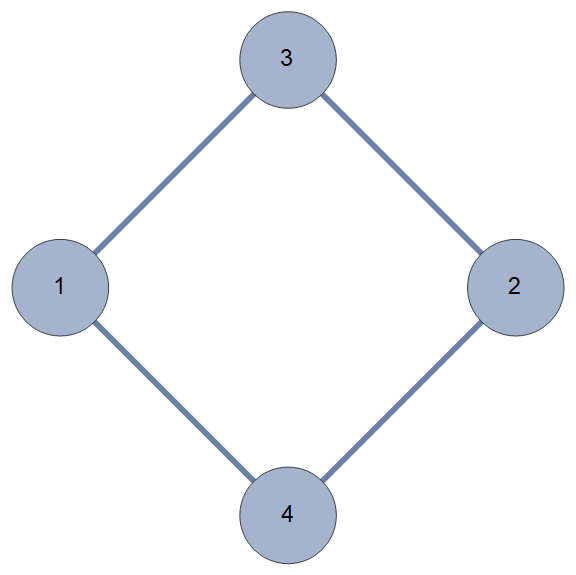}
    \end{minipage}
    \begin{minipage}{0.7\linewidth}
      \centering
      \includegraphics[width=\linewidth]{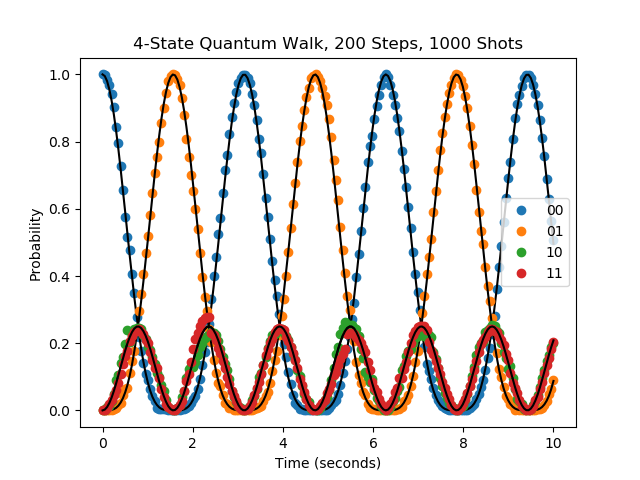}
    \end{minipage}
\end{figure}

\section{Appendix B - Some more 8-vertex quantum walks}

\begin{figure}[H]
    \centering
    \begin{minipage}{0.3\linewidth}
      \centering
      \includegraphics[width=\linewidth]{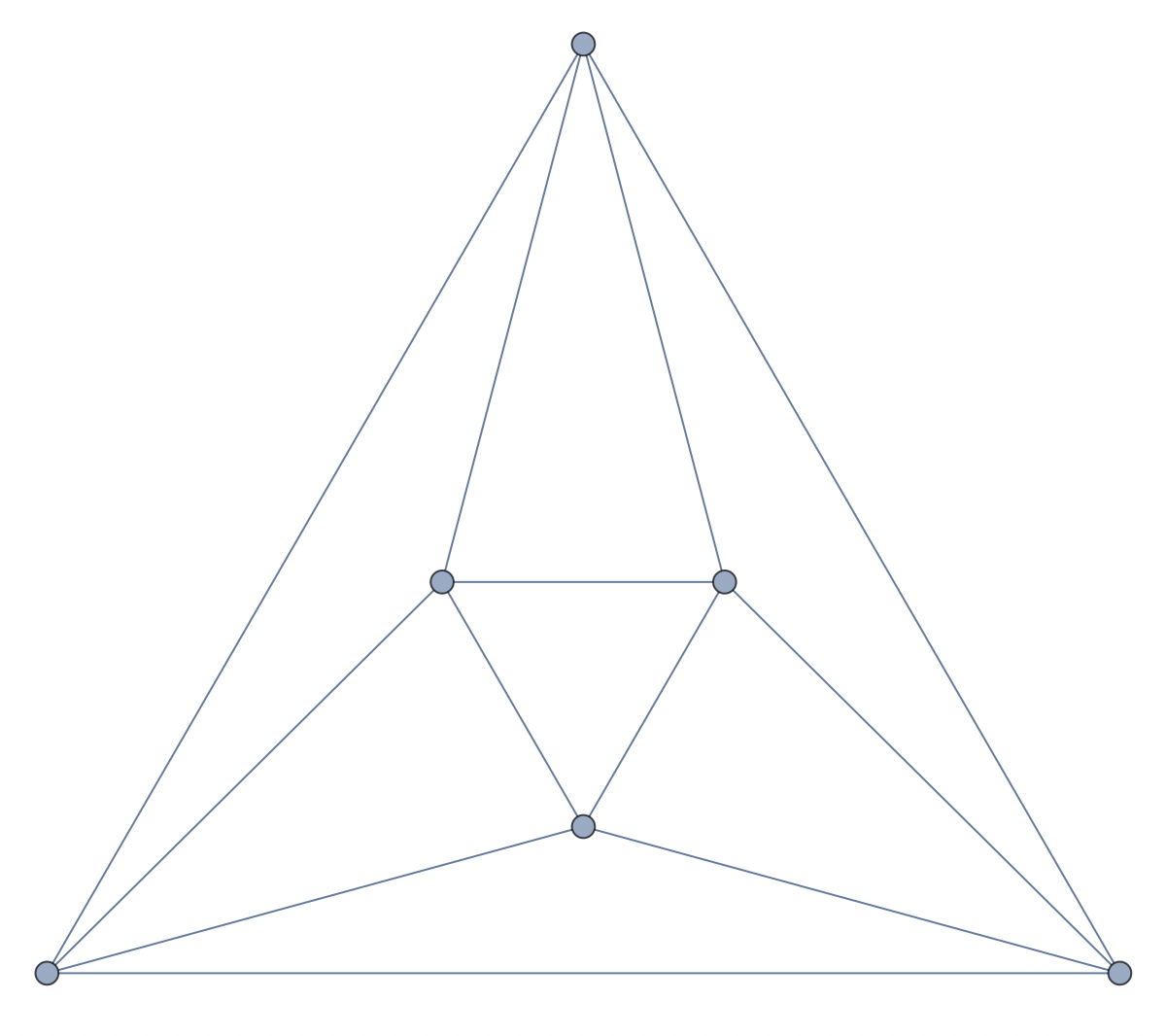}
    \end{minipage}
    \begin{minipage}{0.7\linewidth}
      \centering
      \includegraphics[width=\linewidth]{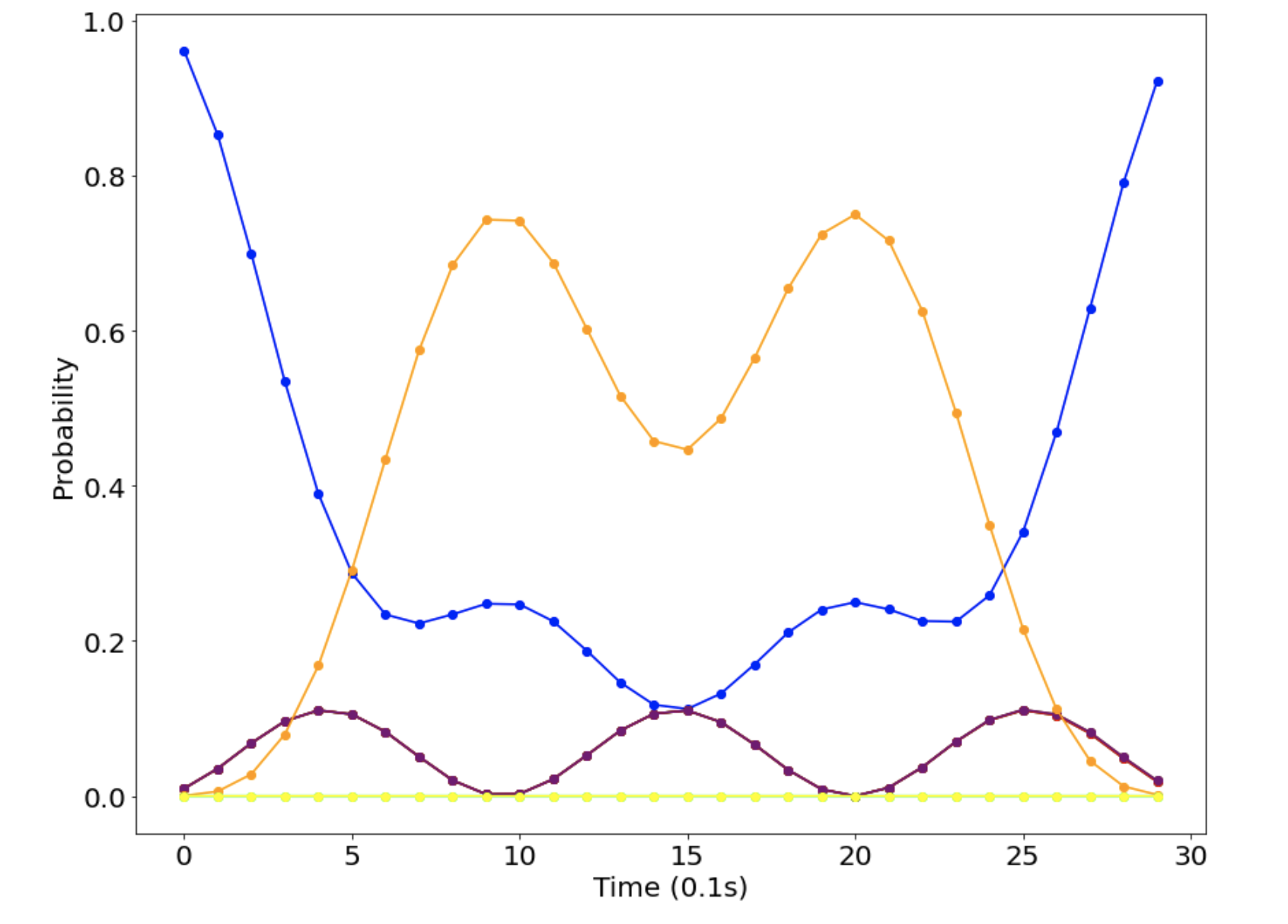}
    \end{minipage}
\end{figure}

\begin{figure}[H]
    \centering
    \begin{minipage}{0.3\linewidth}
      \centering
      \includegraphics[width=\linewidth]{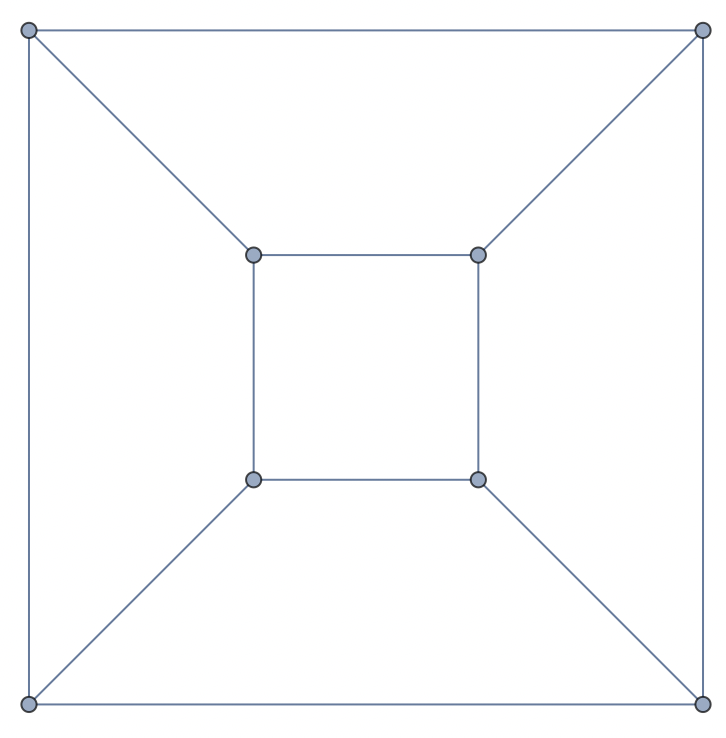}
    \end{minipage}
    \begin{minipage}{0.7\linewidth}
      \centering
      \includegraphics[width=\linewidth]{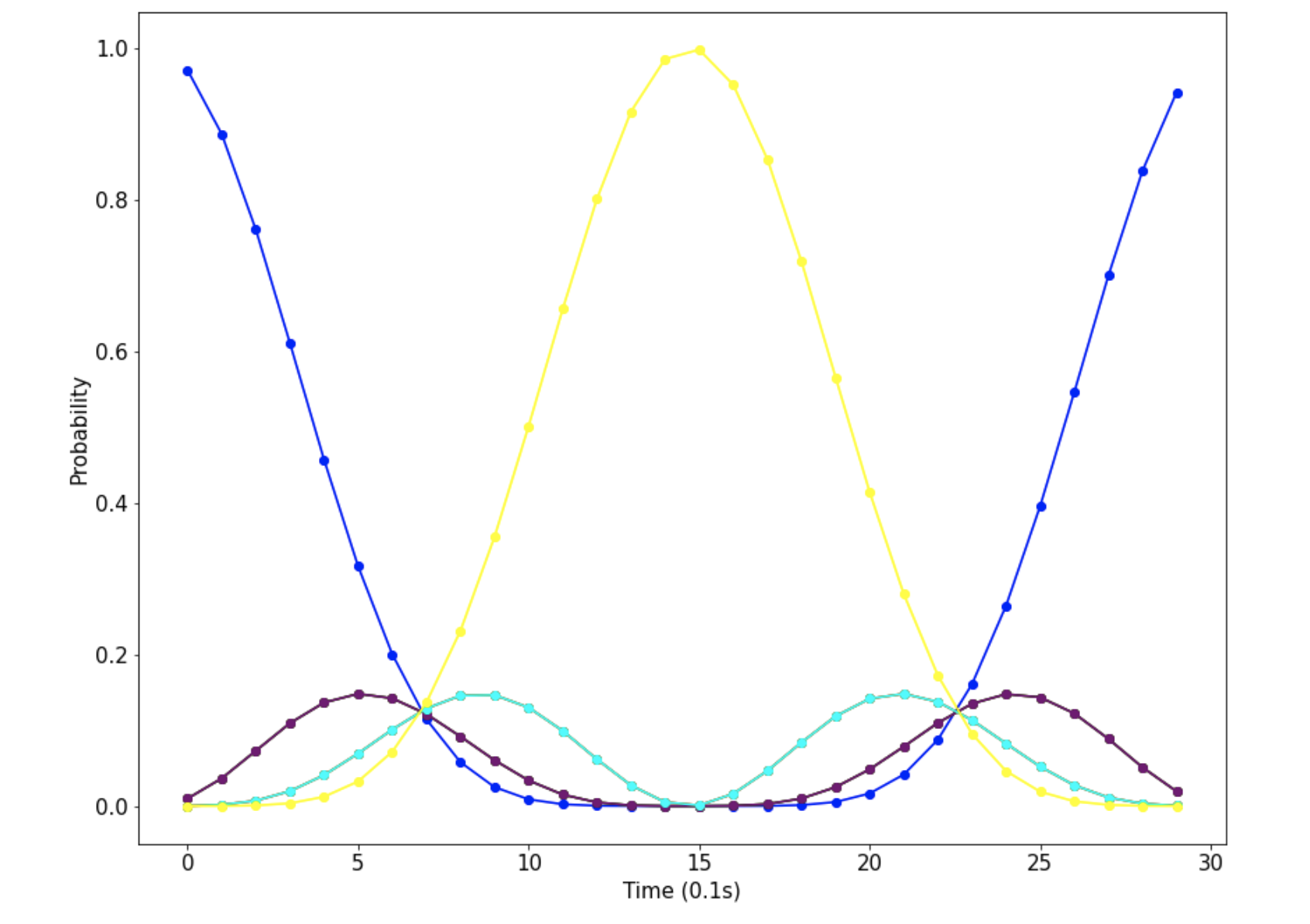}
    \end{minipage}
\end{figure}

\begin{figure}[H]
    \centering
    \begin{minipage}{0.3\linewidth}
      \centering
      \includegraphics[width=\linewidth]{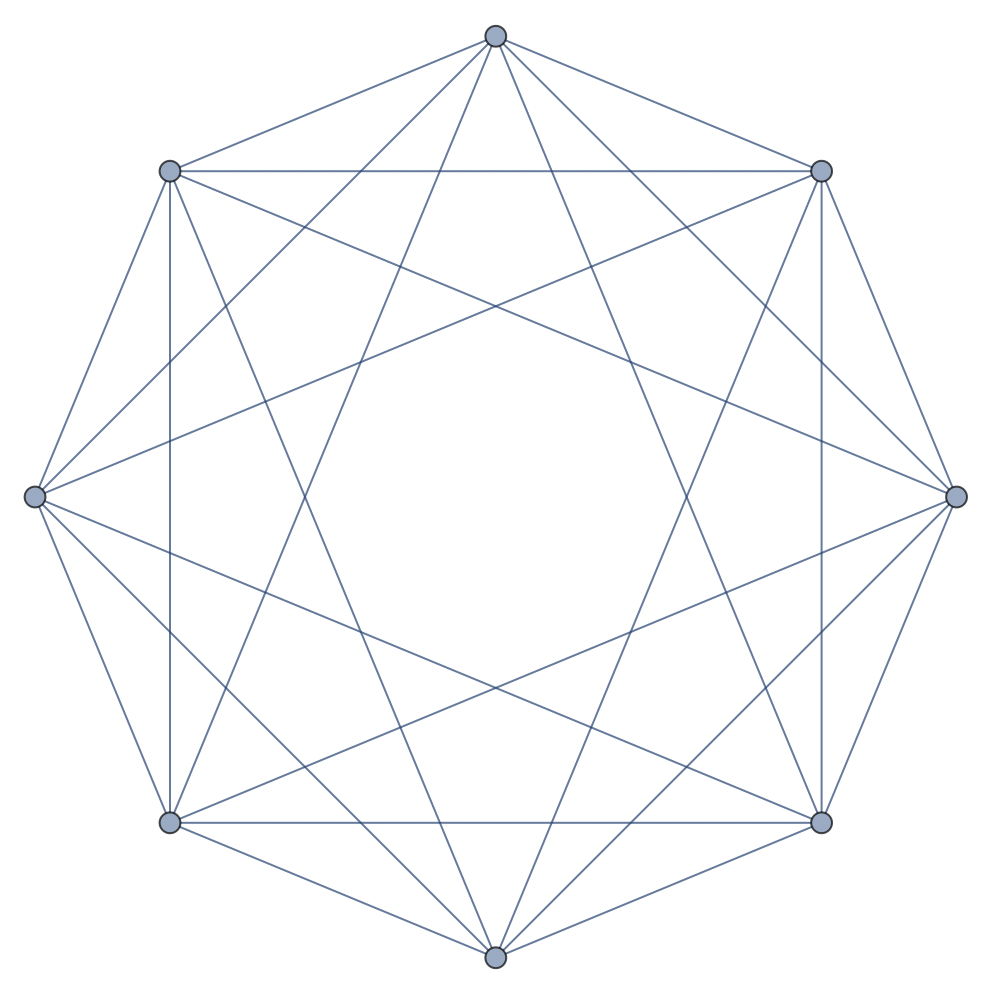}
    \end{minipage}
    \begin{minipage}{0.7\linewidth}
      \centering
      \includegraphics[width=\linewidth]{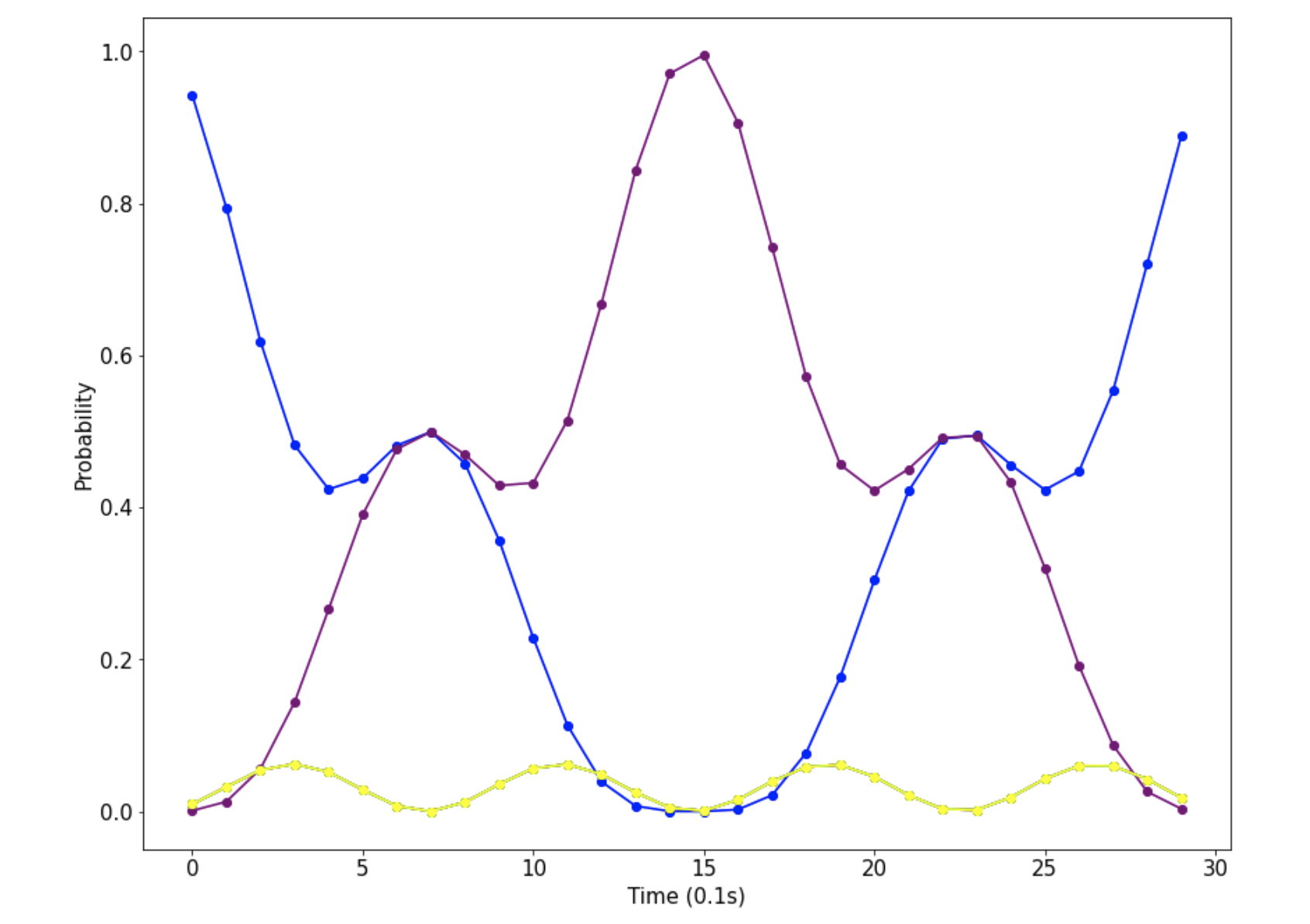}
    \end{minipage}
\end{figure}

\end{document}